\documentclass{article}
\usepackage{latexsym}
\usepackage{amsmath}
\usepackage{epsfig}

\newcommand{\alfa}{\mbox{$\alpha$}}
\newcommand{\av}[1]{\mbox{$ \langle #1 \rangle $}}

\newcommand{\aabha}{\mbox{\tt SAMBHA}}
\newcommand{\BHLUMI}{\mbox{\tt BHLUMI}}
\newcommand{\QQ}{\mbox{$q^2$}}
\newcommand{\Da}{\mbox{$\Delta\alfa$}}
\newcommand{\mzz}{\mbox{$m_Z^2$}}

\newcommand{\ee}{\mbox{$e^+e^-$}}
\newcommand{\proc}[2]{\mbox{$ #1 \rightarrow #2 $}}
\newcommand{\dd}{{\mathrm d}}    
\newcommand{\vecc}[1]{\mbox{\boldmath $#1$}}
\begin{document}
\begin{titlepage}
\title{\bf The running of the electromagnetic coupling $\alpha$ \\
        in small-angle Bhabha scattering}
\author{A.B. Arbuzov
\\
BLTP, Joint Institute for Nuclear Research, Dubna, 141980, Russia 
\\
\\
D. Haidt
\\
DESY, Notkestrasse 85, D-22603 Hamburg, Germany
\\
\\
C. Matteuzzi  M. Paganoni
\\
Dipartimento di Fisica, Universit\`{a} di Milano-Bicocca\\
 and 
\\
INFN-Milano,     Piazza della Scienza 3, I-20126 Milan, Italy
\\
\\
L. Trentadue\thanks{On leave from Dipartimento di Fisica, Universit\`{a} di  Parma and INFN, 
      Gruppo Collegato di Parma, I-43100 Parma, Italy }
\\
Department of Physics, CERN Theory Division, 1211 Geneva 23, Switzerland}
\date{}
\maketitle
\vspace*{2.0cm}
\begin{abstract}
A method to determine the running of \alfa\ from a measurement of 
          small-angle Bhabha scattering is proposed and worked out. The method 
          is suited to high statistics experiments at \ee colliders, which 
          are equipped with luminometers in the appropriate angular region.
          A new simulation code predicting small-angle Bhabha scattering is 
          also presented.
\end{abstract}
\begin{picture}(5,2)(-330,-485)
\put(1.0,-5){} \put(1.0,-15.2){CERN-PH-TH/2004-016}
\put(1.0,-25.4){}
\end{picture}
\vfill
\vspace{3cm}
\thispagestyle{empty}
\end{titlepage}
\section{Introduction}
%
The electroweak Standard Model $SU(2)\otimes U(1)$ contains Quantum Electrodynamics (QED) as a constitutive 
part. The running of the electromagnetic
coupling \alfa\ is determined by the theory as
\begin{equation}
  \alfa(\QQ) = \frac{\alfa(0)}{1-\Da(\QQ)}, \label{eq:Da}
\end{equation}
where \alfa(0)\,=\,$\alpha_0$ is the Sommerfeld
fine structure constant, which has been measured to 
a precision of 3.7$\times$10$^{-9}$\ \cite{pdg00}; $\Delta\alpha(q^2)$
positive arises from loop contributions to
the photon propagator. The numerical prediction of electro\-weak observables
involves the know\-ledge of \alfa(\QQ), usually  for \QQ \,$\neq$\, 0. For 
instance, the know\-ledge of \alfa(\mzz) is relevant 
to the evaluation of quantities 
measured by the LEP experiments. This is achieved by evolving \alfa\ from 
\QQ\,=\,0\ up to the $Z$-mass scale \QQ\, =\, \mzz. 
The evolution expressed by the quantity \Da\ receives contributions from
leptons, hadrons and the gauge bosons. The hadronic contribution to the
vacuum polarization, which 
cannot be calculated from first principles, is estimated with the help of a 
dispersion integral and evaluated \cite{fj} by using total cross section 
measurements of $e^+e^-\rightarrow$ hadrons at low energies. Therefore, any evolved 
value \alfa(\QQ)\, particularly  for $|q^2|>4m_{\pi}^2$, 
is affected by uncertainties 
originating from hadronic contributions.  
The uncertainty on $\alfa(\mzz)^{-1}$ induced by these 
data is as small as $\pm$0.09 \cite{fj}; nevertheless it turned out 
\cite{hhm} that this
limits the accurate prediction of electroweak quantities within the 
Standard Model, parti\-cularly for the prediction of the Higgs mass. 

While waiting for improved measurements from BEPC, VEPP-4M and DAFNE as input
to the dispersion integral, intense efforts are
made to improve on estimating the hadronic shift $\Da_\mathrm{had}$,
as for instance \cite{fj99} - \cite{pp}, and to find alternative ways of measuring
\alfa\ itself. Attempts have been made to measure \alfa(\QQ) directly, using 
\ee data at various energies, such as measuring the ratio
of $e^+e^-\gamma$/$e^+e^-$ \cite{topaz} or more directly the angular 
distribution of Bhabha scattering \cite{L3}.

In this article the running of \alfa\ is studied using small-angle Bhabha 
scattering. This process provides unique information on the QED coupling 
constant \alfa\ at low {\it space-like} momentum transfer $t=-|\QQ|$, where
\begin{equation}
 t = -\frac{1}{2}\ s\ (1-\cos\theta) \label{eq:t}
\end{equation}
is related to the total invariant energy $\sqrt{s}$ and to the scattering 
angle $\theta$ of the final-state electron. The small-angle region 
has the virtue of giving access to values of \alfa($t$)\ without being affected
by weak contributions. The cross section can be theoretically calculated 
with a precision at the per mille level. It is dominated by the photonic
$t$ channel exchange and the non-QED contributions have been
computed~\cite{luca} and are of the order of 10$^{-4}$ (see table~\ref{tab:cross});
in particular, contributions from boxes with two weak bosons are
safely negligible.

In general, the Bhabha cross section is computed (see sect. 3) from the entire set of 
gauge-invariant amplitudes in both the $s$ and $t$ channels. Consequently, two invariant 
scales $s$ and $t$ govern the process. The different amplitudes are functions of both
$s$ and $t$ and also the QED coupling $\alpha$ appears as $\alpha(s)$ resp. $\alpha(t)$
\cite{brodsky}. However, the restriction of Bhabha scattering to the kinematic regime of 
small angles results in a considerable simplification, since 
the $s$ channel then gives only a negligible contribution, as is quantitatively 
demonstrated in table~\ref{tab:cross}. Thus, the measurement of the angular distribution 
allows us indeed to verify directly the running of the coupling $\alpha(t)$. 
For the actual calculations, $\theta\gg m_e/E_\mathrm{beam}$ and 
$E_\mathrm{beam}\gg m_e$ must be satisfied (see sect. 4.1).
Obviously, in order to mani\-fest the running, the experimental precision must be ad\-equate. 

This idea can be realized by high-statistics experiments at $e^+e^-$ colliders
equipped with finely segmented luminometers, in particular by 
the LEP experiments, given their large event samples, by SLC and future Li\-near 
Colliders. The relevant luminometers
cover the $t$-range from a few GeV$^2$ to order 100 GeV$^2$. 

\begin{figure}[h]\centering
   \mbox{\epsfig{file=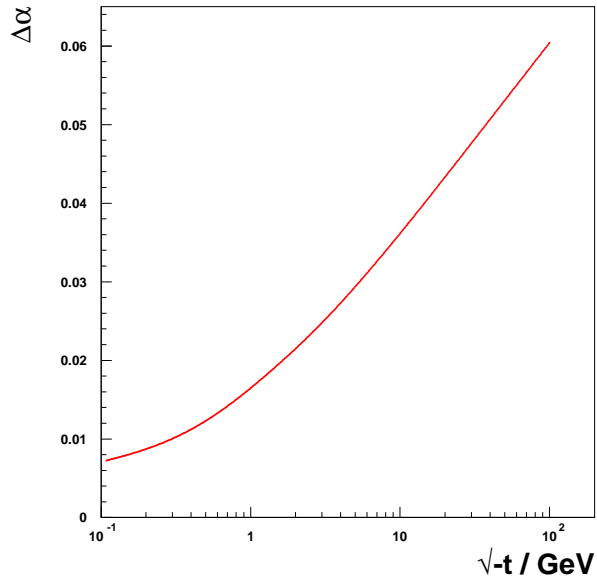,
        bbllx=16,bblly=3,bburx=530,bbury=501,
         width=8.5cm,clip=}}
   \caption{\sl \Da\ versus $\sqrt{-t}$ in units of GeV in the space-like region.}
   \label{fig:alfa}
\end{figure}

\begin{figure}[h]\centering
   \mbox{\epsfig{file=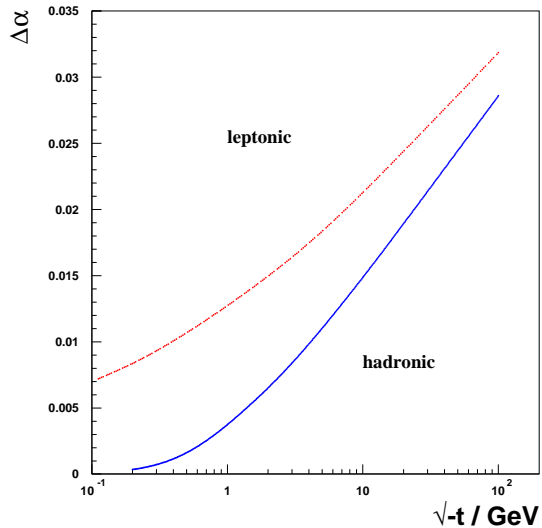,width=8.5cm}}
   \caption{\sl Contributions to \Da\ from leptons (dashed curve) and hadrons 
            (solid curve) versus $\sqrt{-t}$ in units of GeV.}
   \label{fig:alfadet}
\end{figure}

The $t$-dependence of the quantity \Da($t$)\ (eqs.~\ref{eq:Da},~\ref{eq:t}) at small
values of $t$ is illustrated in fig.~\ref{fig:alfa}.
It shows the predicted running of \alfa\ in the relevant space-like region. The 
figure is obtained using the program {\it alphaQED} by Jegerlehner \cite{fj}.
At low energies (see fig.~\ref{fig:alfadet}) \Da\ is dominated by the 
contribution from the leptons,
while with increasing energy also the contribution involving hadrons
gets relevant. The region where hadronic corrections are critical is contained
in the considered $t$-range.
%
\section{The method}
\label{sec:1}
%
The experimental determination of the angular distribution of the 
Bhabha cross section requires the precise de\-finition of a Bhabha event
in the detector. The analysis 
follows closely the procedure adopted in  the luminosity measurement, which 
is described in detail, for instance in ref.~\cite{LEP_report}, and
elaborates on the additional aspect related to the measurement of a 
differential quantity. To this
aim the luminosity detector must have a sufficiently large angular acceptance 
and adequate fine segmentation. The variable $t$ (eq.~\ref{eq:t}) is 
reconstructed on an event-by-event basis. 

The method to measure the running of $\alpha$ exploits the fact that the
cross section for the process \proc{\ee}{\ee} can be conveniently decomposed 
into three factors~:
\begin{equation}
\frac{\dd \sigma}{\dd t} = \frac{\dd \sigma^0}{\dd t}
                     \left(\frac{\alpha(t)}{\alpha(0)}\right)^2
                     (1+\Delta r(t)) \label{eq:meth}
\end{equation}
as worked out in sect. 3.
All three factors are predicted to a precision of 0.1\% or better. 
The first factor on the right-hand side refers to the effective Bhabha Born cross section, 
including soft and virtual photons according to ref.~\cite{luca},
which is precisely known, and accounts for the strongest dependence on $t$. 
The vacuum-polarization effect in the leading photon $t$ channel exchange
is incorporated in the running of $\alpha$ and gives rise to the squared factor in
eq.~\ref{eq:meth}.
The third factor, $\Delta r(t)$, 
collects all the remaining real (in particular collinear) and virtual radiative effects
not incorporated in the running of $\alpha$. 
The experimental data after correction for detector effects have to be compared
with eq.~\ref{eq:meth}. The $t$ dependence is rather steep, 
thus migration effects may need attention.

This goal is achieved by using a newly developed program based on the already existing
semianalytical code {\tt NLLBHA} \cite{luca,nllbha}. A detailed description of this
code called \aabha\ is given in sect.4.

%
\section{Theory}
\label{sec:2}
%
It is convenient to confront the fully corrected measured cross section
with the Bhabha cross section, including radiative corrections in the factorized 
form given by eq.~\ref{eq:meth}. 
The physical cross section is infrared safe \cite{luca}.
This decomposition is neither unique nor dictated by a compelling physical reason;
rather it allows the separation of the different sources of $t$ dependence in
a transparent way without introducing any additional theoretical uncertainty.
The va\-rious factors are discussed one by one in the following subsections.
%
\subsection{The cross section $\dd \sigma^0/\dd t$}
%
The differential cross section $\dd\sigma^0/\dd t$ is defined as~:
\begin{equation}
\frac{\dd \sigma^0}{\dd t} = \frac{\dd \sigma^B}{\dd t}
                       \left(\frac{\alpha(0)}{\alpha(t)}\right)^2.
\end{equation}
The factor $d\sigma^B/dt$ is the Bhabha cross section in the
improved Born approximation, which, by definition, includes the running of 
$\alpha$. As seen explicitly in the formulae below (eq.~\ref{eq:form_B}) 
the term $\alpha(t)/\alpha(0)$ is not factorized completely 
in the improved Born cross section. In order to have the factorized form of
eq.~\ref{eq:meth} the $t$ channel contribution to the
running of $\alpha$ has been taken out. In this way, $\dd \sigma^0/\dd t$
contains not only the usual Born $t$ dependence, i.e. $1/t$, but also some 
weaker $t$ dependences arising from $s$ channel amplitudes with vacuum polarization
effects taken into account~\cite{luca}, although numerically small as mentioned above.

The improved Born cross section for Bhabha scattering within the 
electroweak Standard Model is precisely known (see 
refs.~\cite{Budny,Bardin,Bohm}). The differential cross section $\dd \sigma^B/\dd t$ 
differs from $\dd \sigma^0/\dd t$ by the inclusion of those radiative corrections that
affect only the propagator of the exchanged photon. 
They form a gauge-invariant subset of all radiative corrections and are
shown explicitly.
It is convenient to decompose $\dd \sigma^B/\dd t$ into the contributions arising
from the $t$ channel $(B_t)$, the $s$ channel $(B_s)$ and their interference
$(B_i)$: 
\begin{equation}\label{eq:form_B}
\frac{\dd\sigma^B}{\dd t }\, = \,\frac{\pi \alpha_0^2 }{2s^2} {\mathrm{Re}}
\bigl\{B_t + B_s + B_i \bigr\}, 
\end{equation}
where
\begin{eqnarray*}
B_t &=& \biggl(\frac{s}{t}\biggr)^2 \biggl\{\frac{5+2c+c^2}{(1-\Pi(t))^2} 
    + \xi\frac{2(g_v^2+g_a^2)(5+2c+c^2)}{(1-\Pi(t))} \\  
    &+& \xi^2\biggl(4(g_v^2+g_a^2)^2 
    + (1+c)^2(g_v^4+g_a^4+6g_v^2g_a^2)\biggr)\biggr\} \\
B_s &=& \frac{2(1+c^2)}{|1-\Pi(s)|^2}
    + 2\chi\frac{(1-c)^2(g_v^2-g_a^2) 
    + (1+c)^2(g_v^2+g_a^2)}{1-\Pi(s)} \\
    &+& \chi^2\bigl[(1-c)^2(g_v^2-g_a^2)^2
    + (1+c)^2(g_v^4+g_a^4+6g_v^2g_a^2)\bigr] \\
B_i &=& 2\frac{s}{t} (1+c)^2 \biggl\{ \frac{1}{(1-\Pi(t))(1-\Pi(s))} \\
    &+& (g_v^2+g_a^2)\biggl(\frac{\xi}{1-\Pi(s)}+\frac{\chi}{1-\Pi(t)}\biggr) \\ 
    &+& (g_v^4+6g_v^2g_a^2+g_a^4)\xi\chi\biggr\}
\end{eqnarray*}
\begin{eqnarray*}
\chi &=& \frac{s}{s-m^2_z+im_Z \Gamma_Z}\cdot\frac{1}{\sin2\theta_w}\\
\xi  &=& \frac{t}{t-m^2_Z}\cdot\frac{1}{\sin2\theta_w}\, ,\\
g_a &=& -\frac{1}{2},\quad g_v=-\frac{1}{2}+2 \sin^2\theta_w), \\
t &=& (p_1-q_1)^2=-\frac{1}{2}\;s\;(1-c),\\
c &=& \cos\theta ,\qquad \theta =\widehat{\vecc{p}_1 \vecc{q}_1}.
\end{eqnarray*}
Here $s$ is the total squared invariant mass, $\theta_w$ the electroweak 
mixing angle and $\theta$ the scattering angle between the initial and final electron
with momentum $\vecc{p}_1$ and $\vecc{q}_1$ respectively (see ref.~\cite{luca}).

In table~\ref{tab:cross} the cross sections are given in nanobarns  for 
the pure QED and electroweak cases. QED$_t$ denotes the contribution 
of the $t$ channel pure QED Feynman diagrams. The cross sections are integrated over
two relevant angular ranges.
\begin{table*}[ht]
\caption{Various cross sections in {\rm nb} as a function of the centre-of-
         mass energy in {\rm GeV} integrated over the two angular ranges 
         45--110 mrad and 5--50 mrad. 
         The index $t$ denotes the contribution of the
         corresponding $t$ channel Feynman diagrams alone.
         The last columns are of interest for furture Linear Colliders.
}
\begin{center}
\begin{tabular}{|c|c|c|c|c|c|c|c|}
\hline
$\sqrt{s}$ ({\rm GeV})& 91.187 & 91.2     & 189     & 206     & 500
& 1000    & 3000     \\
\hline
& \multicolumn{7}{c|}{45 mrad $\ <\ \theta \ <\ $ 110 mrad} \\ \hline
$\sqrt{\av{-t}}$ (GeV) & 3.4 &  3.4 & 7.1 & 7.7 & 18.8 & 37.5    & 112.6 \\ \hline
QED               & 51.428     & 51.413   & 11.971  & 10.077  & 1.7105
& 0.42763 & 0.047514 \\
QED$_t$           & 51.484     & 51.469   & 11.984  & 10.088  & 1.7124
& 0.42809 & 0.047566 \\
EW                & 51.436     & 51.413   & 11.965  & 10.072  & 1.7105
& 0.42871 & 0.049507 \\
EW+VP$_t$         & 54.041     & 54.018   & 12.743  & 10.745  & 1.8590
& 0.47303 & 0.055748 \\
EW+VP             & 54.036     & 54.013   & 12.742  & 10.744  & 1.8588
& 0.47296 & 0.055742 \\
\hline
& \multicolumn{7}{c|}{5 mrad $\ <\ \theta \ <\ $ 50 mrad} \\ \hline
$\sqrt{\av{-t}}$ (GeV) & 1.1 & 1.1 & 2.2 & 2.4 & 5.8 & 11.6 & 34.8\\ \hline
QED               & 4963.4     & 4962.0   & 1155.4  & 972.54  & 165.08
& 41.271  & 4.5857   \\
QED$_t$           & 4963.5     & 4962.1   & 1155.4  & 972.57  & 165.09
& 41.272  & 4.5858   \\
EW                & 4963.4     & 4962.0   & 1155.4  & 972.53  & 165.08
& 41.272  & 4.5885   \\
EW+VP$_t$         & 5075.0     & 5073.5   & 1190.6  & 1003.3  & 172.51
& 43.647  & 4.9603   \\
EW+VP             & 5075.0     & 5073.5   & 1190.6  & 1003.3  & 172.51
& 43.646  & 4.9605   \\
\hline
\end{tabular}
    \label{tab:cross}
\end{center}
\end{table*}
The table shows that the $t$ channel photon exchange
dominates the cross section at small angles and justifies why the
process is suited for investigating the $t$ dependence, and so the running
of $\alpha(t)$.

By comparing the values of the electroweak cross section with
the pure QED one, it is  seen that the $Z$-boson exchange
gives a negligible contribution to small-angle scattering. In the last 
two lines (EW+VP$_t$ and EW+VP) there are numbers for the cross section with
vacuum polarization taken into account in the $t$ channel only, and 
in all channels, correspondingly. One can see that the effect of $s$ channel 
vacuum polarization is small, as a result of the smallness of the $s$ channel 
photon--exchange contribution itself. 
The last line in the table corresponds to the complete formula in 
eq.~\ref{eq:form_B}.

%
\subsection{The running of $\alpha$}
%
In eq.~\ref{eq:form_B} the two-point functions $\Pi(t)=\Delta\alpha(t)$ and 
$\Pi(s)=\Delta\alpha(s)$ are responsible for the running of $\alpha$
in the space-like and time-like regions. In the language of Feynman diagrams 
the effect arises from fermion-loop insertions into the
virtual photon lines:
\begin{eqnarray*}
\Pi(t) &=& \frac{\alpha_0}{\pi} \;\biggl(\delta_t
        +\frac{1}{3}L-\frac{5}{9}  \biggr) \\
       &+&\biggl(\frac{\alpha_0}{\pi}\biggr)^2\biggl(\frac{1}{4}\;L + \zeta(3)
        - \frac{5}{24} \biggr) \nonumber \\   
       &+& \biggl(\frac{\alpha_0}{\pi}\biggr)^3\Pi^{(3)}(t)+ {\mathcal O}\biggl(\frac{m_e^2}{t}\biggr),
\end{eqnarray*}
where
\begin{eqnarray*}
L=\ln\frac{Q^2}{m_e^2},\qquad  Q^2=-t,\qquad \zeta(3)=1.202
\end{eqnarray*}
and where the leading part of the two-loop contribution to the polarization 
operator is taken into account. 
The most significant part arises from the electrons and is $L/3-5/9$.

The ${\mathcal O}(\alpha)$ and
${\mathcal O}(\alpha^2)$ leptonic vacuum polarization has been known for many
years~\cite{Kallen}. The third-order (three--loop) leptonic contributions 
$\Pi^{(3)}(t)$ have recently been calculated~\cite{Steinhauser}.
In the Standard Model, $\delta_t$ contains contributions from muons, 
$\tau$-leptons, $W$-bosons and hadrons~:
\begin{eqnarray*}
 \delta_t &=& \delta_t^{\mu} + \delta_t^{\tau} + \delta_t^{W}
           +\delta_t^H, \\ 
 \delta_s &=&\delta_t\;(t\rightarrow s), 
\end{eqnarray*}
which means that $\delta_s$ is obtained from $\delta_t$ by substituting
$s$ by $t$, see ref.~\cite{luca}.
The contributions from the leptons ($l$\,=\,$\mu,\tau$) and from the $W$
are theoretically calculable and given by:
\begin{eqnarray*}
 \delta_t^{l} &=& \frac{1}{2}\;v_{l}\;\biggl(1-\frac{1}{3}v^2_{l}\biggr)
 \;\ln\frac{v_{\l}+1}{v_{l}-1}+\frac{1}{3}\;v^2_{l}
 - \frac{8}{9} \\
 v_{l} &=&\sqrt{1+\frac{4m_{l}^2}{Q^2}}\, , \\
 \delta_t^{W} &=& \frac{1}{4}\;v_{W}\;(v^2_{W}-4)
  \;\ln\frac{v_{W}+1}{v_{W}-1}-\frac{1}{2}\;v^2_{W}
  +\frac{11}{6} \\
  v_W &=&\sqrt{1+\frac{4M_{W}^2}{Q^2}}\, .   \nonumber
\end{eqnarray*}
For $Q^2\gg m_l^2$ the formula simplifies to 
\begin{eqnarray*}
 \delta_t^{l}  &=& \frac{1}{3}\ln\frac{Q^2}{m_{l}^2}-\frac{5}{9}, 
 \nonumber
\end{eqnarray*}

The hadronic contribution cannot be calculated theoretically; instead,
it can be expressed as a dispersion integral involving experimentally 
measured $e^+e^-$ cross sections:
\begin{eqnarray}
&& \delta_t^\mathrm{had}=\frac{Q^2}{4\pi\alpha_0^2}\int\limits_{4m_{\pi}^2}^{\infty}
\frac{\sigma^{e^+e^-\rightarrow \mathrm{h}}\;(s')}{s'+Q^2}\;\dd s'.
\label{eq:disp}
\end{eqnarray}
For numerical calculations, hadronic contributions as included in
the parametrisation of refs. ~\cite{fj99,Eidelman} are adopted.

This procedure, as usually assumed (see e.g. \cite{dav}),
is based on the analyticity of the function $\alpha(q^2)$ in the complex plane, except possibly at the
energies corresponding to the Landau pole. For the leptonic contributions
$\delta_{s,t}^{e,\mu,\tau}$ this assertion is true, while for the hadronic 
contribution $\delta_t^\mathrm{had}$ it relies on the dispersion approach to
the entire, non-perturbative, hadronic physics (see eq.~\ref{eq:disp}).
This ends up in a single analytical function that can be used to deal with the
vacuum polarization in the $t$ channel.

\subsection{The radiative factor $1+\Delta r(t)$ and neglected terms} 

For the present investigation of the small-angle Bhabha cross section
only the corrections consistently needed to maintain the required accuracy are kept. 
All these corrections are included in the new code \aabha. All the following 
contributions have been proved to be negligible \cite{luca} and are dropped~:
\begin{itemize}
\item Any electroweak effect beyond the tree level, for instance appearing 
      in boxes or vertices with $Z^0$ and $W$ bosons, running weak coupling, etc.
\item Box diagrams at order $\alpha^2$ and larger
\item Contributions of order $\alpha^2$ without large logarithms, leading from 
      order $\alpha^4$  (i.e. $\alpha^4 L^4$, $\cdots$) and subleading higher
      order ($\alpha^3 L^2$, $\alpha^4 L^3$, $\cdots$)
\item Contributions from pair-produced hadrons, muons, taus and the corresponding
      virtual pair corrections to the vertices (estimated to be of the order
      of 0.5$\times$10$^{-4}$).
\end{itemize}

The radiatively corrected Bhabha cross section is denoted by $\dd \sigma/\dd t$. Numerically 
it differs from  $\dd \sigma^B/\dd t$ by less than a few per cent for small angles, depending
on energy and final-state selection procedure.

%
\section{Monte Carlo codes and comparison}
%
 
The precise determination of the luminosity at $e^+e^-$ colliders is a crucial
ingredient to obtain an accurate evaluation of all the physically relevant
cross sections. They necessa\-rily have to rely on some reference
process, which is usually taken to be the small-angle Bhabha scattering.
Given the high
statistical precision provided by the LEP collider, an equally precise
know\-ledge of the theoretical small-angle Bhabha cross section is mandatory.
In the 1990's the substantial progress in measuring the
luminosity reached
by the LEP machine has prompted several groups to make a theoretical
effort aiming at a 0.1\% accuracy~\cite{LEP_report,Arbuzov}. This goal has 
indeed been achieved by developing a dedicated strategy.
For the first time small-angle Bhabha scattering was evaluated analytically,
following a new calculation technique~\cite{luca} that yields the required
precision. Analytical calculations have been
combined with Monte Carlo programs in order to simulate realistically
the conditions of the LEP experiments.

The analytical results evaluated for the various contributions
to the observed Bhabha cross section in ref.~\cite{luca},
were implemented into the semi--analytical code
{\tt NLLBHA} (for a short write--up see in ref.~\cite{nllbha}).
The important feature of this code consists in the systematic account of
all QED radiative corrections required to reach the {\it per mille}
precision. On the other hand, the simulation of realistic experimental
acceptances can only be achieved with Monte Carlo techniques. For this purpose
a Monte Carlo code, {\tt LABSMC}, was developed \cite{labsmca,labsmcb,labsmcc}.

\subsection{\aabha -{\tt NLLBHA}}

The program {\tt LABSMC}, which was intended to describe large-angle Bhabha scattering
at high energies,  has been complemented
with a set of routines from {\tt NLLBHA} so as to be
applicable to small-angle Bhabha 
scattering. This implied the insertion of the relevant second-order
next--to--leading radiative corrections $({\mathcal O}(\alpha^2L))$ in the 
Monte Carlo code\footnote{The codes are available upon request from the
authors.}, which are crucial to achieve the per mille accuracy. The extension
to cover small angles resulted in the new code \aabha\, containing the 
previously existing features together with the following new characteristics :
\begin{itemize}
\item the complete electroweak matrix element at the Born level;
\item the complete set of ${\mathcal O}(\alpha)$ QED radiative corrections
      (including radiation from amplitudes with $Z$-boson exchange);
\item vacuum-polarization corrections by leptons,
      hadrons~\cite{Eidelman}, and $W$-bosons;
\item 1--loop electroweak radiative corrections and effective EW couplings
      by means of the {\tt DIZET v.6.30}~\cite{dizet} package;
\item higher-order leading-logarithm photonic corrections by means of the
      electron structure functions~\cite{Kuraev,Nicrosini,Skrzypek,Arb};
\item light pair corrections in the ${\mathcal O}(\alpha^2L^2)$ 
      leading-logarithm approximation including (optionally) the
      two-photon and singlet mechanisms.
\end{itemize}
The code is applicable with the following restrictions:
\begin{itemize}
\item[a)] $E_{\mathrm{beam}}\gg m_e$: the energy has to be much larger 
          than the electron mass;
\item[b)] $m_e/E_{\mathrm{beam}} \ll \theta$: extemely small
          angles are not described well, but the condition is fulfilled 
          in practice for both small-  and large-angle Bhabha 
          measurements in the experiments at LEP, SLC and NLC;
\item[c)] starting from the second order in $\alpha$,
          real photon emission is integrated over, i.e. events with two
          photons se\-parated from electrons are not generated.
\end{itemize}
%
 
\subsection{{\BHLUMI}}
 
The Monte Carlo Program \BHLUMI\, which has been used in the LEP analyses,
is described in detail in ref.~\cite{jad97}.
 
\subsection{Comparison between {\BHLUMI} and \aabha}

\BHLUMI\ is compared with \aabha\ for integral and, for the first time, also
differential distributions. The actual measurements are
of calorimetric type. Therefore, event samples are generated with both
programs, subjecting each event to a common set of calorimeter-like  criteria
(hereafter called CALO).
 
In a first test the program codes were applied to the conditions established
by the working group {\it Event gene\-rators for Bhabha scattering } \cite{LEP_report}, 
with the result that all numbers were reproduced within the quoted accuracy.

\begin{figure}[h]
   \centering
   \mbox{\epsfig{file=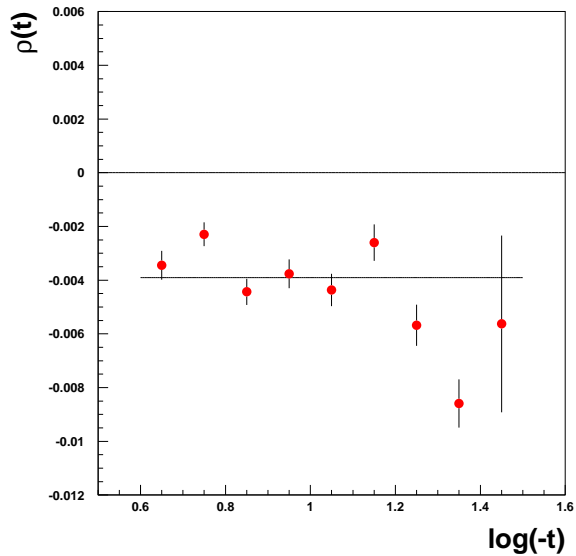,
         bbllx=18,bblly=15,bburx=529,bbury=500,clip=,
        width=8cm}}
   \caption{\sl Cross section ratio $\rho(t)$ as a function of
log ($-t$), with $t$ in units of GeV$^2$ .}
   \label{fig:ratio_calo}
\end{figure}

In a further test, about 10$^8$ Bhabha events were ge\-nerated according to 
the calorimeter like conditions spe\-cified in sect. 5.1. 
This selection rejects a considerable part of events with real hard photon 
radiation. Therefore, the effect of mutual
cancellation between virtual and real radiation is suppressed, which 
inevitably causes fairly large $t$-dependent radiative corrections.
The comparison is made quantitative in the form of the ratio
\begin{eqnarray*}
 \rho(t) = \frac{\dd \sigma_\mathrm{sambha}/\dd t-\dd \sigma_\mathrm{bhlumi}/
\dd t}
      {\dd \sigma_\mathrm{bhlumi}/\dd t}
\end{eqnarray*}
and displayed in fig.~\ref{fig:ratio_calo}. A linear logarithmic fit to the cross-section ratios and their statistical 
uncertainties gives 
\begin{eqnarray*}
  \rho(t) = -(0.0039\pm 0.0002) -(0.0046\pm 0.0010) \ {\rm log}\frac{-t}{\av{-t}}
\end{eqnarray*}
with $\av{t}=\,-\, 8.3 GeV^2$.

The two programs differ significantly, on average by 0.4\%. At the present level
of investigation, it cannot be excluded that there is a weak $t$ dependence.

It is not so surprising to find a discrepancy for the differential quantity, while
getting good agreement for the integral quantity. In fact, the key element is the
far stronger restriction in the event selection for the two cases. In the integral case
the events are accepted over the entire angular range of the luminometer, while for the
differential analysis the same selection criteria are applied to a set of segments
covering eventually the full range of the luminometer. This implies that an event 
accepted in the integral case is not necessarily accepted in the differential one
owing to the more restrictive conditions, so that the sum of events accepted in the
segmented luminometer is smaller than the number of events in the full luminometer.

\begin{table*}[ht]
\caption{Comparison between the codes {\tt NLLBHA} (\aabha) and \BHLUMI. 
Numbers are obtained by using conditions on table 19 in ref.~\cite{LEP_report}.
The relative ratio  $\delta r/r$ in per mille is defined by (YR-NOW)/YR.
Last column gives the relative difference between \BHLUMI (NOW) and 
\aabha(NOW).}

\begin{center}
\begin{tabular}{|l|ccr|ccr|r|}
\hline
cut & BHLUMI (YR) & BHLUMI (NOW) & $\delta$r/r &  NLLBHA (YR) & NLLBHA (NOW) & $\delta$r/r & \\ \hline
0.1  &  166.892  &  166.879  &  0.07  &  166.948  &  166.923 &  0.14  & $-$0.26   \\
0.3  &  165.374  &  165.438  & $-$0.38  &  165.448  &  165.420 &  0.16  &  0.10   \\
0.5  &  162.530  &  162.616  & $-$0.52  &  162.561  &  162.25  &  1.91  &  2.25   \\
0.7  &  155.668  &  155.733  & $-$0.41  &  155.607  &  155.40  &  1.33  &  2.13   \\
0.9  &  137.342  &  137.425  & $-$0.60  &  137.199  &  137.32  & $-$0.88  &  0.76   \\
\hline
\end{tabular}
\label{tab:accordo}
\end{center}
\end{table*}

For a quantitative understanding of this qualitative argument, a selection of events
is presented as a function of the cut $s\cdot x_c=s\cdot x_1 x_2$, where  $x_i$
is the fractional energy carried by the electron (or positron) 
(see  table~\ref{tab:accordo}). 
Obviously, a value
for $x_c$ near to 1 selects configurations with small acollinearity, as opposed to
cases with smaller $x_c$, which favour larger acollinearity configurations. For a given
opening angle, events with large acollinearity are hardly accepted; in other words
the size of the cone opening  angle defines the number of radiative events
containing real emitted photons accepted or rejected. Consequently a larger or smaller
final-state phase space is probed. Since virtual radiative contributions are unaffected
by phase-space restrictions, the interplay between real and virtual 
radiative contributions strongly depends on the acceptance. The accuracy to which
radiative corrections have to be treated becomes crucial.

With the tight cuts required for the study of a differential quantity, as in the
case investigated here, fine detailed aspects related to radiative
contributions are  necessarily probed. Therefore such studies open a new level of comparison between
theory and experiment. 

%
\section{Evaluation of the running in a simulated experiment}

Anticipating the application of the proposed method to measure the $t$ dependence of
$\alpha(t)$ on the data of a real experiment, a Monte Carlo simulation is carried out 
instead, in order to demonstrate the feasibility. An event sample is generated in the 
conditions of the 
DELPHI experiment using the existing program \BHLUMI. In the next subsection the sample
 so obtained is confronted
with the expectation of the new program \aabha. It should be noted that the $t$ dependence
of $\alpha(t)$, i.e. the quantity to be investigated, is stronger by about an order of magnitude
than the possible differences in the intrinsic $t$ dependences between \BHLUMI\
and \aabha  (see sect.4.3).

%
\subsection{Event generation}
%
The DELPHI detector and its performance are described in ref.~\cite{del}.
For the analysis, the relevant subdetector is the electromagnetic calorimeter STIC~\cite{del},
which covers the extreme forward and backward directions. It has a ring structure with 
segmentation in both $\theta$ and $\phi$ covering $\sqrt{-t}$ ranges from 1.5 to 6~GeV 
for LEP1 energies and 3 to 12 GeV for LEP2 energies. 

Electrons, positrons and photons are observed as clusters. Their reconstruction is 
based on a cluster algorithm. The Bhabha events are characterized by two 
narrow high energy electromagnetic clusters opposite to each other and 
well inside the detector. The cluster algorithm is applied to the observed 
energy depositions in the cells of the electromagnetic calorimeter.
Furthermore, the cluster with the highest energy satifies the more restrictive 
requirement to be at the radial position $R$ between 10 and 25 cm such as to cause 
no inefficiency for the opposite cluster.

A Monte Carlo simulation has been performed using \BHLUMI\ for three centre-of-mass 
energies of LEP: 91.2 ($Z$ peak), 189 and 200 GeV. Assuming 
integrated luminosities $\int{\cal{L}}\dd t$ typical of the LEP experiments, 
the number of events passing the selection criteria is obtained and listed 
in table~\ref{tab:sample}. An event is attributed to ring $i$, if the highest
energy cluster is reconstructed in this ring and  the criteria listed below
are satisfied.
\begin{itemize}
\item Cluster reconstruction:\\
      The main criterion for merging adjacent cells is:
\begin{equation}
\left( \frac{\Delta\theta}{\,30\, \rm mrad}\right)^2+
\left( \frac{\Delta\phi}{870\, \rm \, mrad}\right)^2 < 1 \nonumber
\end{equation}
      where the cluster centre is calculated as the energy-weighted cell
      centres.
 
\item Only the highest energy cluster in each hemi\-sphere (referred to
      as {\it F(forward)} and {\it B(backward)}\,) is considered
 
\item Energy requirements:\\
      min($E_F$,$E_B$) $>$ 0.65 $E_\mathrm{beam}$ \\
      max($E_F$,$E_B$) $>$ 0.94 $E_\mathrm{beam}$ \\
      This implies that the Bhabha events have not suffered from sizeable
      initial-state radiation effects.
 
\item Geometrical acceptance: \\
      The radial position $R$ of the two opposite clusters must satisfy \\
\begin{equation}
 7 {\,\rm cm} < R_F,R_B < 28 {\,\rm cm} \nonumber
\end{equation}
\item Kinematics:\\
      The cluster centre and the nominal interaction point of the colliding
      \ee\ beams determine the dip angle $\theta$. The quantity
      $t$ is calculated from the dip angle $\theta$ and the
      nominal centre-of-mass energy $\sqrt{s}$ = 2$E_\mathrm{beam}$ according to
\begin{equation}
  t = -\frac{1}{2}s(1-\cos\theta_\mathrm{max})\nonumber
\end{equation}
      where $\theta_\mathrm{max}$ is defined to be the dip angle of the cluster with the 
      highest energy.
\end{itemize}
The result of the Monte Carlo experiment is summarized in table~\ref{tab:sample}.
Ring 1 and ring 7 have been disregarded in order to exclude any inefficiency from 
border effects. 
\begin{table}[h]
\caption{Numbers of events generated with \BHLUMI}
\begin{center}
\begin{tabular}{|l|rrr|}
\hline
$\sqrt{s}$(GeV) & 91.2 & 189 & 200 \\ \hline
$\int{\cal{L}}\dd t$ (pb$^{-1}$) & 75 & 150 & 200 \\ \hline
Ring 2   & 1844850 & 863571 & 1028210 \\ 
Ring 3   &  907754 & 425586 &  506131 \\ 
Ring 4   &  513696 & 240550 &  286994 \\ 
Ring 5   &  313218 & 146731 &  174740 \\ 
Ring 6   &  201893 &  94033 &  112168 \\ 
\hline
\end{tabular}
\label{tab:sample}
\end{center}
\end{table}
%

%
\subsection{Comparison and evaluation}
%

In this subsection the relevant observables and the para\-meters to be extracted
are established and discussed. 

Each ring defines with its boundaries a bin ($t_\mathrm{min},t_\mathrm{max}$). The event numbers
are to be equated to the corresponding theoretical prediction obtained from the
formulae implemented in the program \aabha. 
In order to extract the $t$ dependence of $\alpha(t)$, eq.~\ref{eq:meth}
is evaluated for each ring $R_i$ defined by the geometry of the DELPHI luminometer. 
Equation~\ref{eq:meth} then reads, for ring $i$~:
\begin{equation}
  \sigma_i = \sigma^0_i 
             \biggl(\frac{\alpha(t_i)}{\alpha(0)}\biggr)^2
              (1+\Delta r_i),
  \label{eq:meth_i}
\end{equation}
with the following definitions~:
\begin{eqnarray*}
 \sigma_i &=& \int^{R_i} \dd t \frac{\dd \sigma}{\dd t}
\\ 
 \sigma^0_i &=& \int^{R_i} \dd t \frac{\dd \sigma^0}{\dd t}
\\ 
 \biggl(\frac{\alpha(t_i)}{\alpha(0)}\biggr)^2  &=&
 \int^{R_i} \frac{\dd t}{t_{\mathrm{max}}-t_{\mathrm{min}}}
 \biggl(\frac{\alpha(t)}{\alpha(0)}\biggr)^2,
\\ 
 1+\Delta r_i &=&  \biggl(\frac{\alpha(0)}{\alpha(t_i)}\biggr)^2
                    \frac{\sigma_i}{\sigma^0_i}
 \label{eq:ave}
\end{eqnarray*}
Table~\ref{tab:theo} contains the resulting theoretical values.
%
\begin{table*}[ht]
\caption{Theoretical predictions for each ring of the three factors
         of eq.~\ref{eq:meth_i}. For the conditions defined in sect. 5.1
         the angular boundary of ring $i$ is 
         $\theta_i$=arctan (7+3(i-1))/220). 
}
\begin{center}
\begin{tabular}[]{|c|c|c|c|c|c|c|c|}
\hline
No. of ring  & 1 & 2 & 3 & 4 & 5 & 6 & 7 \\ \hline
& \multicolumn{7}{c|}{$\sqrt{s}=91.2$ GeV} \\ \hline
$\sigma^0_i$
& 63.077 & 24.728 & 12.170 & 6.8694 & 4.2517 & 2.8120 & 1.9552  \\
$\biggl(\alpha(t_i)/\alpha(0)\biggr)^2$
& 1.0425 & 1.0475 & 1.0516 & 1.0551 & 1.0582 & 1.0609 & 1.0634 \\
$1+\Delta r_i$
& 0.9426 & 0.9440 & 0.9412 & 0.9395 & 0.9240 & 0.8915 & 0.7982 \\
\hline
& \multicolumn{7}{c|}{$\sqrt{s}=189$ GeV} \\ \hline
$\sigma^0_i$
& 14.685 & 5.7563 & 2.8324 & 1.5984 & 0.9889 & 0.6537 & 0.4542 \\
$\biggl(\alpha(t_i)/\alpha(0)\biggr)^2$
& 1.0554 & 1.0613 & 1.0661 & 1.0702 & 1.0736 & 1.0767 & 1.0794 \\
$1+\Delta r_i$
& 0.9377 & 0.9390 & 0.9360 & 0.9329 & 0.9165 & 0.8858 & 0.7898 \\
\hline
& \multicolumn{7}{c|}{$\sqrt{s}=200$ GeV} \\ \hline
$\sigma^0_i$
& 13.115 & 5.1406 & 2.5295 & 1.4274 & 0.8831 & 0.5838 & 0.4057 \\
$\biggl(\alpha(t_i)/\alpha(0)\biggr)^2$
& 1.0565 & 1.0625 & 1.0673 & 1.0714 & 1.0749 & 1.0780 & 1.0807 \\
$1+\Delta r_i$
& 0.9376 & 0.9387 & 0.9352 & 0.9330 & 0.9158 & 0.8847 & 0.7896 \\
\hline
& \multicolumn{7}{c|}{$\sqrt{s}=1000$ GeV} \\ \hline
$\sigma^0_i$
& 0.5248 & 0.2059 & 0.1014 & 0.0573 & 0.0356 & 0.0236 & 0.0165 \\
$\biggl(\alpha(t_i)/\alpha(0)\biggr)^2$
& 1.0921 & 1.0994 & 1.1050 & 1.1096 & 1.1135 & 1.1169 & 1.1199 \\
$1+\Delta r_i$
& 0.8622 & 0.8620 & 0.8590 & 0.8545 & 0.8398 & 0.8084 & 0.7205 \\
\hline
& \multicolumn{7}{c|}{$\sqrt{s}=3000$ GeV} \\ \hline
$\sigma^0_i$
& 0.0590 & 0.0234 & 0.0117 & 0.0067 & 0.0042 & 0.0028 & 0.0020 \\
$\biggl(\alpha(t_i)/\alpha(0)\biggr)^2$
& 1.1192 & 1.1267 & 1.1325 & 1.1373 & 1.1414 & 1.1448 & 1.1479 \\
$1+\Delta r_i$
& 0.8467 & 0.8457 & 0.8422 & 0.8381 & 0.8253 & 0.7956 & 0.6975 \\
\hline
\end{tabular}
\label{tab:theo}
\end{center}
\end{table*}
%

Putting together the experimental and theoretical ingredients, i.e.
the observed number of events $N_i$ in each ring, together with
the relevant luminosities $\int{\cal{L}}\dd t$ (from table~\ref{tab:sample}) and
$\sigma_i^0$, $\Delta r_i$ (from table~\ref{tab:theo}), we obtain the final formula:
\begin{equation}
 \biggl(\frac{\alpha(t_i)}{\alpha(0)}\biggr)^2 =
     \frac{N_i}{{\sigma^0_i\int{\cal{L}}\dd t}}\frac{1}{1+\Delta r_i},
 \label{eq:run}
\end{equation}
which can be exploited in a linear fit to access the para\-meters defining the
$t$ dependence of $\alpha$~:
\begin{equation}
 \biggl(\frac{\alpha(t)}{\alpha(0)}\biggr)^2 =
   (u_0\pm\delta u_0) + (u_1\pm\delta u_1)\cdot {\rm log}\frac{-t}{\av{-t}}
 \label{eq:u_exp}
\end{equation}
The parameters of the fit are listed in table~\ref{tab:fits}.
\begin{table}[h]
\caption{Table of fit results; the uncertainties $\delta u_0$ and $\delta u_1$
         are uncorrelated.}
\begin{center}
\begin{tabular}{|c|ccc|}
\hline
$\sqrt{s}$ & 91.2 GeV & 189 GeV & 200 GeV \\ \hline
$u_0$   & 1.0573$\pm$0.0005   & 1.0698$\pm$0.0008 & 1.0703$\pm$0.0007 \\ \hline
$u_1$   & 0.0242$\pm$0.0028   & 0.0284$\pm$0.0041 & 0.0318$\pm$0.0038 \\ \hline
\av{-t} & 8.5 GeV$^2$         & 36.6  GeV$^2$     & 40.9   GeV$^2$    \\ \hline
\hline  
\end{tabular}
\label{tab:fits}
\end{center}
\end{table}
%
\section{Discussion}

Table~\ref{tab:fits} demonstrates that for the case of the DELPHI
setup (see sect.~5) and assuming typical integrated luminosities,
the statistical accuracy is sufficient to verify the running of
$\alpha$ for each of the three centre-of-mass energies. 

Equation~\ref{eq:run} can be expanded in terms of $\Delta\alpha$ (see 
eq.~\ref{eq:Da}). It is convenient to consider 
\begin{equation}
 \frac{N_i}{\sigma^0_i}\frac{1}{1+\Delta r_i} = n_0+n_1
                       \log\frac{-t_i}{\av{-t}}
 \label{eq:exp}
\end{equation}
rather than eq.~\ref{eq:run}, since in practice the integrated luminosity 
$\int{\cal{L}}\dd t$ is not known. 
The two coefficients $n_0$ and $n_1$ are obtained from a linear fit
and contain the information on both  the data and theory. 
Their interpretation is :
\begin{eqnarray*}
 n_0 &=& \int{\cal{L}}\dd t\cdot \biggl(1+2\Delta\alpha(\av{t})\biggr) \\
 n_1 &=& \int{\cal{L}}\dd t\cdot \biggl(\frac{\dd}{\dd \log (-t)}2\Delta\alpha(t)\biggr) .
\label{eq:result}
\end{eqnarray*}
The dependence on the integrated luminosity is given explicitly: obviously,
one has $n_i$ = $u_i\cdot \int{\cal{L}}\dd t$ by comparing eqs.~\ref{eq:run},
\ref{eq:u_exp},\ref{eq:exp}. 

In the ratio $n_1/n_0$  the dependence of the integrated luminosity drops out :
\begin{eqnarray*}
 \frac{\dd}{\dd \log (-t)}\Delta\alpha &=&
                \frac{n_1}{2n_0}\biggl(1+2\Delta\alpha(\av{t})\biggr)
\end{eqnarray*}
The slope $\dd \Delta\alpha/\dd \log(-t)$, the quantity of interest, is then directly
given by the ratio of the two experimentally measured quantities $n_0$ and $n_1$, 
namely $n_1/2n_0$. The contribution of $2\Delta\alpha(\av{t})$ is small
with respect to 1 and can be neglected. The accuracy of the slope is determined by 
$\delta n_1/2n_0$, i.e. about 10\% (see table~\ref{tab:fits}), which is far smaller 
than the absolute value of  $n_1/2n_0$. 

On the other hand,  $n_0$ relates the integrated luminosity to
$\Delta\alpha$ at the average value of $t$ 
\begin{equation}
  \int{\cal{L}}\dd t = \frac{n_0}{1+2\Delta\alpha(\av{t})}
   \nonumber
\end{equation}
Making use of $\Delta\alpha(\av{t})$ as a priori knowledge
the fitted $n_0$ can be used
to derive the integrated lumino\-sity, which is the standard procedure. The
statistical precision is given by $\delta n_0/n_0$, which is of the order of 10$^{-3}$.

In addition, the hadronic contribution to $\Da(t)$ (see fig.~\ref{fig:alfadet})
may be deduced by subtracting the leptonic contribution, which is theoretically 
known precisely. The extraction of the hadronic contribution is only limited by the
experimental precision.
%
\section{Conclusions}
%
A novel approach to access directly and to measure the running of 
$\alpha$ in the space-like region
is proposed. It consists in analy\-sing small-angle Bhabha scattering.
Depending on the particular angular detector coverage and on the energy of the
beams, it allows a sizeable range of the $t$ variable to be covered.

The feasibility of the method has been put in evidence by the use of a new
tool, \aabha\ , to calculate the small-angle Bhabha differential cross section with 
a theoretical accuracy of better than 0.1\%.

The information obtained in the $t$ channel can be compared with the existing 
results of the $s$ channel measurements. This represents a complementary approach,
which is direct, transparent and based only on QED interactions and furthermore
free of some of the drawbacks inherent in the $s$ channel methods. 

The method outlined can be readily applied to the experiments at LEP and SLC.
It can also be exploited by future \ee\ colliders as well as by existing lower
energy machines.

An extremely precise measurement of the QED running coupling $\Delta\alpha(t)$ 
for small values of $t$ may be envisaged with a dedicated luminometer 
even at low machine energies.
%
\section*{Acknowledgements}
%
We are grateful to the  CERN and DESY directorates and to the Universities of Parma 
and Milan for the hospitality and the support extended to us during the course of this 
work. Two of us (A.A. and L.T.) want to acknowledge the INTAS Organization for support 
at early stages of this work. One of us (L.T.) wants to thank the INFN and the Italian 
Ministry for University and Scientific and Technological Research (MURST) for
financial support. 
We are indebted to Marco Incagli, Graziano Venanzoni and Carlos Wagner for the invitation
to present part of these results at the WIN02 Conference and the Sighad03 Workshop.  
We enjoyed a fruitful discussion with P.M.~Zerwas.

\newpage

\end{document}